\documentstyle[aps,prl,twocolumn]{revtex}
\begin{document}
\def\B.#1{{\bf#1}}
\title{Comment on ``Multicomponent turbulence, the spherical limit,\\
and non-Kolmogorov spectra''} 
\author{Victor L'vov, Evgenii Podivilov$^*$  and Itamar Procaccia} 
\address{Department of Chemical Physics, 
The Weizmann Institute of Science, Rehovot 76100, Israel,\\
$^*$ Institute of Automation and Electrometry, Ac. Sci. of Russia, 630090,
Novosibirsk, Russia}
\maketitle
\begin{abstract}
 It is shown that the generalization of the Navier-Stokes equations to
 a theory with $N$ ``internal state" copies of the velocity fields is a
 step in a wrong direction: the $N\to\infty$ limit has no physical 
 sense and produces wrong results, whereas the treatment of the first
 order terms in $1/N$ is even more complicated than  the initial problem
 of description of turbulence in the frame of the Navier-Stokes equation.
\end{abstract}

Consider an incompressible velocity field ${\bf u}({\bf r},t)$ which
is the solution of the Navier Stokes equations
\begin{equation}
 \partial {\bf u}/ \partial t + ({\bf u}\cdot {\bbox{\nabla}})
  {\bf u} -\nu\nabla^2 {\bf u} +{\bbox{\nabla }} p =0 \ , \quad
  {\bbox{\nabla}}\cdot {\bf u}=0\,, 
\label{NS} 
\end{equation} 
where $\nu$ is the kinematic viscosity  and $p$ is the pressure. We
  assume that there exist appropriate boundary conditions to maintain
  a high Reynolds number flow.  It has been suggested in the
  literature, and especially recently\cite{95MW}, that it is
  advantageous to consider a generalization of this equation to a
  situation in which there are $N$ copies of the velocity field
  labeled by an ``internal" index $s$:
\begin{equation}
\B.u(\B.r,t) \Rightarrow \B.u_s(\B.r,t), 
\quad s=1,2,\dots,N \ . 
\label{copy}
\end{equation}
In terms of these copies one writes the generalized equation for
incompressible fields  ${\bf u}_s$:
\begin{equation}
 \partial {\bf u}_s/  \partial t + A_{slm}({\bf u}_l\cdot
  {\bbox{\nabla}}) {\bf u}_m -\nu\nabla^2 {\bf u}_s +{\bbox{\nabla }}
  p_s = 0\ .
\label{NScopy} 
\end{equation} 
The hope is that the statistical properties of this theory are
simpler to elucidate than those of (\ref{NS}) in the limit
$N \to \infty$. We will show here that this hope is not realized when
the proper symmetries are taken into account.

Consider first the implication of Galilean invariance. Our theory has
to remain invariant to the transformation
\begin{equation}
\B.r \to \B.r' \equiv \B.r -\B.U_0 t\,,\ \  
\B.u_s \to \B.u'_s \equiv \B.u_s+h_s\cdot \B.U_0 \ . \label{G1}
\end{equation}
In (\ref{G1}) $h_s$ is a scalar in 3-space, but a vector in the
internal state space. In case when the velocity fields $\B.u_s$ have a
physical sense (like in the two-fluid models of plasma) Galilean
invariance requires $h_s=1$ for every $s$. However our future
conclusions are independent of the particular choice of $h_s$, which
may be considered as a free parameter.

Applying the transformation (\ref{G1}), to (\ref{NScopy}) we find two
additional terms. The time derivative leads to an extra term which
$-(\B.U_0\cdot \nabla)\B.u_s$, whereas the nonlinear term leads to the
extra term $A_{slm}h_l(\B.U_0\cdot \nabla)\B.u_m)$.  Since these two
terms must cancel, we find the constraint on $A_{slm}$ which follows
from the fundamental symmetry of hydrodynamics -- Galilean invariance:
\begin{equation}
A_{slm}h_l = \delta_{sm} \ .
\label{constraint1}
\end{equation}

Next remind that in the known cases of simplification of a problem in
the large $N$ limit (see e.g.\cite{73Abe,73Ma}) one needs
to have some continuous group of symmetry, the Lie group. Important
example of the group Lie are the group of rotations in $3$ dimensional
space SO(3) and the group of unitary matrixes $2\times2$ SU(2).
Elements of the Lie groups $g$ may be parametrizied by a set of
continuous parameters, like the Euler angles for the SO(3)
group. Following\cite{95MW} and aiming possibility of
$1/N$-simplification we equip the copy space with a symmetry group Lie
$\cal G$. By analogy with the theories\cite{73Abe,73Ma} the index $s$
have to be considered as the label of the basis of the representations
of $\cal G$.  Denote by $\hat T(g)$ the operator that corresponds to
the the element $g$ of the group, and the transformed field by $\tilde
\B.u_s$. Then 
\begin{equation} \tilde \B.u_s \equiv \big(\hat T(g)
\B.u \big)_s = T_{ss'}(g)\B.u_s' \ , 
\label{trans} 
\end{equation}
 where $N\times N$ matrix $T_{ss'}(g)$ is the representation of $\hat
T(g)$ in this basis. Now we apply $\hat T(g)$ to the equations of
motion (\ref{NScopy}) and demand invariance. Write for that these
equations for $\B.u_{s'}$ and multiply $T_{ss'}(g)$ from the left. Then
\FL
\begin{equation}
 \! {\partial \tilde {\bf u}_s \over \partial t}
 + T_{ss'}(g)A_{s'lm}({\bf u}_l\! \cdot \! 
  {\bbox{\nabla}}) {\bf u}_m -\nu\nabla^2 \tilde{\bf u}_s 
 +{\bbox{\nabla }} \tilde p_s = 0 . 
\label{NStrans}
\end{equation}
Next we use the fact that $\cal G$ is a group and therefore the matrix
$T_{ss'}$ has an inverse $ \B.u_s = T^{-1}_{ss'}(g) \tilde \B.u_{s'}$.
Substituting in (\ref{NStrans}) and demanding invariance leads to
the constraint
\begin{equation}
T_{ss'}(g)A_{s'l'm'}T^{-1}_{l'l}(g)T^{-1}_{m'm}(g)=A_{slm}
\ . \label{constraint2}
\end{equation}
This is the constraint on $A_{slm}$ which follows from the fact that
the set of Eqs. (\ref{NScopy}) is invariant with respect to the
transformation (\ref{trans}) of the group $\cal G$.
 
Now we have two constraints (\ref{constraint1}) and
(\ref{constraint2}) on the same tensor $A_{slm}$ which will be
considered as a restriction on the allowed form of the transformation
(\ref{trans}) of the group $\cal G$. To find this restriction we
multiply the left hand side of (\ref{constraint2}) by $h_l$ and sum up
on $l$. Together with (\ref{constraint1}) it gives:
\begin{equation}
T_{ss'}(g)A_{s'l'm'}T^{-1}_{l'l}(g)T^{-1}_{m'm}(g)h_l = \delta_{sm}
\ . \label{bang}
\end{equation}
This equation is simplified, by multiplying on the left by $T^{-1}$
and on the right by $T$ to obtain $A_{sml'}T^{-1}_{l'l}(g)h_l
 =\delta_{sm}$.  It appears now that if we have uncountably many
constraints (since $g$ is continuous) on the finite dimensional tensor
$A_{sml}$. The only way to remove the over determination is to demand
that $T^{-1}_{l'l}(g)$ is $g$-independent. Because for the identical
transformation ($g=e$) $T_{ss'}(e) = \delta_{ss'}$ we have
\begin{equation}
T_{ss'}(g) = \delta_{ss'} \ . \label{diag}
\end{equation}
Finally, if we apply (\ref{dial})  to (\ref{constraint2}) we find that
the equipment of the copy space with a continuous symmetry group
leaves $A_{slm}$ without any additional constraint. Equation
(\ref{constraint2}) becomes an empty identity. The way to understand
this startling result is that the requirement of Galilean invariance
introduced an anisotropic ray $h_s$ in the copy space, and there does
not exist a nontrivial transformation that leaves it invariant. In
other words the requirement of the $\cal G$-symmetry of the set of
Eqs. (\ref{NScopy}) itself gives nothing because the whole problem
includes the Galilean invariance (\ref{G1}) which contradicts to this
symmetry.

The point to understand now is that the conclusion that the
coefficients $A_{slm}$ are unconstrained eliminates any hoped for
advantages of the $1/N$ expansion. The technical remark is that in
successful applications of this method the one-loop diagrams have an
$N$ weight that is larger than that of two or more-loop diagrams. The
way this works in practice\cite{95MW} is through some ``coherency''
conditions in $A_{slm}$ that cause a 2-loop diagram to count $N$
less than in a product of two 1-loop diagrams. However, if $A_{slm}$
is unconstrained by the choice of the symmetry group, there is no loss
of $N$ factors in the 2-loop diagrams compared with the 1-loop
diagrams, and no simplification of the diagrammatics appears in the $N
\to \infty$ limit.

Yet in a recent contribution there was an attempt to overcome this
hopeless situation by a clever trick. The idea of \cite{95MW} is
to consider the $N$-dimensional space (\ref{copy}) of additional
unphysical vector fields $\B.u_s (\B.r ,t)$ in which Galilean
invariance (\ref{G1}) is {\em not respected}, coupled to
one physical velocity field $\B.u_0 (\B.r ,t)$ for which Galilean
invariance is {\em retained}. The result of this attempt was a
prediction that the scaling exponent $\zeta_2$ of the second order
structure function in the $N\!\to \! \infty$ limit attains the value of
$\zeta_{2,N\to\infty}=1/2=0.5$ which differs from the experimental
value $\zeta_{2,{\rm exp}}\simeq 0.70$ (see e.g. \cite{zeta2}) much
large than the well known Kolmogorov 1941 prediction $\zeta_{2,{\small
\rm K41}}=2/3 \simeq 0.67$. There is an immediate reason for worry
about this prediction. It has been observed many years ago by
Kraichnan\cite{59Kra} that disregarding the Galilean symmetry and
truncating the diagrammatic at a finite order produces a $1/2$
prediction for  $\zeta_2$. Is it possible that the result
reported in\cite{95MW} is essentially identical to that? We think
so. In fact, on page 3761 of\cite{95MW} one finds the following
observations: ``{\sl Since we have seen that the zero mode} [the
physical velocity $\B.u_0 (\B.r ,t)$] {\it contributes negligible to
the internal bonds of a graph when $N\to \infty$, we may now
completely neglect the zero mode in resuming these diagrams.}  This
means that the theory for the $N$ unphysical modes in the $N\to\infty$
limit is decoupled from the physical velocity field and becomes a
theory in which Galilean invariance has been totally discarded. Then
we find ``{\em In particular, we shall now see that the Green's
function and the double correlator} [of the physical velocity $\B.u_0
(\B.r ,t)$] {\em may be expressed completely in terms of the Green's
function and double correlator} [of the unphysical modes $\B.u_s (\B.r
,t)$]. It follows from the possibility to neglect (for $N\to\infty$)
the own nonlinearity of of the physical velocity with respect to $N$
nonlinear contributions of unphysical modes. Thus the scaling behavior
of the physical velocity is totally determined by the unphysical
modes.  The conclusion is that the Galilean invariance disappears from
the problem in the formulation of Ref.\cite{95MW} in $N\to\infty$
limit.

Clearly for $N=0$ we have the initial Navier-Stokes based formulation
of the hydrodynamic turbulence. Therefore one may hope to have
something reasonable\cite{95MW} in the next terms of $1/N$ expansion
when some tail of the Galilean invariance will recover. On the page
3745 of \cite{95MW} we found: ``{\em A real test of our approach would
be to compute the first correction in powers of $1/N$\dots}''. This
corrections originate from not only two-loops diagrams of unphysical
fields but also from ALL ORDERS diagrams with respect to own
nonlinearity of the physical velocity field. There are no $1/N$
simplification in the later series. Therefore the task to find the
$1/N$ corrections is equivalent to solution the problem of the
Navier-Stokes hydrodynamic turbulence. 

In summary, it was suggested the multicomponent generalization of the
problem of turbulence such that the zero-order approximation is
solvable. However the next step occurs to be as complicated as the
whole solution of the initial problem. It is not unexpected and
happens always when the zero-order problem has no relation to the
initial one. In considered case this is so because the Galilean
invariance (broken in zero-order) as stressed years ago by Kraichnan,
is crucial for hydrodynamics. In short, the multicomponent extension
of the velocity field discussed in\cite{95MW} is a step in a  wrong
direction which makes the following steps even more complicated.


\begin{references}
\bibitem{95MW} C.-Y. Mou and P.B. Weichman, Phys.~Rev.~E {\bf 52}, 3738 (1995)
\bibitem{73Abe} R. Abe, Prog. Theor. Phys. {\bf 49} 113, 1074, 1877 (1973).
\bibitem{73Ma}  S.-K. Ma, Phys.  Rev. A {\bf 7} 2172 (1973).
\bibitem{zeta2} F. Anselmet, Y. Gagne, E.J. Hopfinger, and R.A. Antonia, 
        J. Fluid Mech. {\bf 140}, 63 (1984); R. Benzi, S. Ciliberto,
        R. Tripiccione, C. Baudet, F. Massaioli and S. Succi,
        Phys. Rev. E {\bf 48}, R29 (1993).
\bibitem{59Kra} R. H. Kraichnan, J. Fluid Mech. {\bf 5}, 497 (1959). 

\end{references}
\end{document}